\documentclass[Journal,NoLineNumbers,letterpaper,InsideFigs]{ascelike-new}

\usepackage[displaymath]{lineno}
\usepackage{lmodern}

\usepackage{graphicx} 
\usepackage{amsmath}
\usepackage{newtxtext,newtxmath}
\usepackage[colorlinks]{hyperref}
\AtBeginDocument{%
  \hypersetup{%
    linkcolor=blue,%
    citecolor=black,%
  }%
}%
\usepackage{inconsolata}
\usepackage{listings}
\usepackage[frozencache,cachedir=.]{minted}
\usepackage{enumitem}
\usepackage{comment}
\usepackage{xcolor}
\setminted[fortran]{
frame=lines,
framesep=2mm,
baselinestretch=1.2,
fontsize=\footnotesize,
linenos}
\definecolor{codegreen}{rgb}{0,0.6,0}
\definecolor{codegray}{rgb}{0.5,0.5,0.5}
\definecolor{codepurple}{rgb}{0.58,0,0.82}
\definecolor{backcolour}{rgb}{0.95,0.95,0.92}
\usepackage{booktabs}
\usepackage{float}
\usepackage{tikz}
\usepackage{tabularx}
\usetikzlibrary{shapes.geometric, arrows}
\tikzstyle{process} = [rectangle, minimum width=3cm, minimum height=1cm, text centered, draw=black] 

\usepackage{subcaption}
\usepackage{algorithm}
\usepackage[noEnd=false]{algpseudocodex}

\newcommand{\ds}              { \displaystyle }
\newcommand{\bfm}[1]             { \mathbf{#1}     }
\newcommand{\uu}              { \bfm{u} }

\newcommand{\ww} {\bfm{w}}

\newcommand\dt[1]{\frac{\partial #1}{\partial t}}

\newcommand\dx[1]{\frac{\partial #1}{\partial x}}
\newcommand\dy[1]{\frac{\partial #1}{\partial y}}
\newcommand\area[2]{\left(#1,#2\right)_{\Omega_e}}
\newcommand\edge[2]{\langle #1, #2 \rangle_{\partial \Omega_e}}

\title{GPU-acceleration of the Discontinuous Galerkin Shallow Water Equations Model (DG-SWEM)  with OpenACC}
\author[1]{Chayanon Wichitrnithed}
\author[2,3]{Eirik Valseth} 
\author[4]{Ethan J. Kubatko}
\author[5,6]{Shintaro Bunya}
\author[1]{Clint Dawson}

\affil[1]{Oden Institute for Computational Engineering and Sciences, The University of Texas at Austin. Email: namo@utexas.edu}
\affil[2]{The Norwegian University of Life Sciences}
\affil[3]{Simula Research Laboratory}
\affil[4]{The Department of Civil, Environmental, and Geodetic Engineering, The Ohio State University}
\affil[5]{Coastal Resilience Center, The University of North Carolina at Chapel Hill}
\affil[6]{Institute of Marine Sciences, The University of North Carolina at Chapel Hill}

\NameTag{Wichitrnithed, \today}

\begin{document}

\maketitle
\begin{abstract}
This paper presents a porting of {DG-SWEM}, a discontinuous Galerkin solver for hurricane storm surge, to NVIDIA GPUs. Time-explicit discontinuous Galerkin methods contain a large number of degrees of freedom but have been shown to exhibit a large amount of data parallelism due to the loose coupling between elements, and thus are naturally mapped to the GPU architecture. A previous framework in porting DG-SWEM to GPUs required converting subroutines from Fortran to C++ to be used with CUDA C++. By using OpenACC and Unified Memory, we simplify the porting process and maintain a single codebase for both CPU and GPU versions.
We test the code using a large Hurricane Harvey scenario on NVIDIA's Grace Hopper chip, and compare the GPU code's performance on multiple H200 nodes to the CPU version on the same amount of Grace CPU nodes.
\end{abstract}

\section{Introduction}
One of the primary causes of flooding on the Texas and US Gulf and Atlantic
coasts is storm surge: the inundation of coastal regions by water in the ocean that is driven
inland by hurricanes or tropical storms. This phenomenon is extremely devastating to life
and property of those nearshore as well as coastal ecosystems. 
In mitigating damage and
guiding relief and emergency efforts, computer simulation of storm surge is critical in both
operational forecasting as well as in hindcasting past storms for validation, engineering planning, and design studies.
Some well-established and commonly used models include, for storm surge, the ADvanced CIRCulation (ADCIRC) model~\cite{Pringle2020,luettich1992adcirc},  the Sea, Lake, and Overland Surges from Hurricanes (SLOSH)~\cite{jelesnianski1992slosh}, and Delft3D~\cite{veeramony2017forecasting}; they have all been extensively validated for past hurricane events. 

In this paper, the  focus is on a model closely related to ADCIRC, the Discontinuous Galerkin Shallow Water Equation Model (DG-SWEM) \cite{kubatko2006hp,dawson2011discontinuous,Wichitrnithed2024-jc}  and aims to improve the main criticism of this model, i.e., the computational cost.
The formulation of ADCIRC is based on the two-dimensional shallow water equations (SWE) which are obtained by depth-averaging the incompressible Navier-Stokes equations.
The advantage of using finite elements is the ability to model complex coastal domains
through unstructured meshes. At the same time, a straightforward usage of the continuous Bubnov-Galerkin (CG) method  results in spurious oscillations for advection-dominated flows. ADCIRC handles this by using a reformulation of the continuity equation in the SWE called the Generalized Wave Continuity Equation (GWCE), but as a consequence of this and the CG methodology, local  mass conservation is
no longer guaranteed. Furthermore, in practice, it still necessary to turn the advective terms in the GWCE off for some cases to achieve numerical stability. 

One critical type of flooding that presents difficulties for ADCIRC is compound flooding
which arises when multiple water sources such as storm surge and rainfall with resulting river discharge
occur simultaneously. 
In practice, ADCIRC has encountered instability when rainfall is added to the source term of the GWCE and its subsequent interaction with wetting and drying scheme.
DG-SWEM has recently~\cite{Wichitrnithed2024-jc} been shown to be able to address many of these problems by using
the discontinuous Galerkin (DG) finite element formulation on the primitive SWE \cite{dawson2011discontinuous,Dawson2002-wv}. 
The DG method can be viewed as a synthesis of the Galerkin method and the finite volume method, whereby the solution is approximated using polynomial basis functions local to each finite element (cell) which guarantees local conservation. Solutions need not have a single value at each node (vertex) and coupling between adjacent elements is done through a numerical flux. It has been widely applied to hyperbolic conservation laws in various domains such as electromagnetics, acoustics, plasma physics, and shallow water equations. We refer to \cite{Hesthaven2008-on,Cockburn1998-su} for the history and details of the DG method.

A primary disadvantage of the DG method which has confined DG-SWEM to mainly research and not operational forecasting like ADCIRC is the increased
computational cost associated with the computations required from the higher number of  global degrees of freedom compared to the continuous formulation. At the same time, the localized nature of computing the solution for each element independently generates a high degree of data parallelism, and several GPU implementations of the DG method have been using different approaches. Early implementations using NVIDIA's CUDA include \cite{Klockner2009-is,Siebenborn2012-ew,Fuhry2013-xg,brodtkorb2012efficient}. More recent implementations using different application programming interfaces (API) can be seen in \cite{Abdi2019-rs,Kirby2020-al} which use OCCA, and \cite{Xia2014-um,Dai2022-jf} which use OpenACC. 
 Lastly, we mention the recent open-source FEniCSx SWE solverby Dawson \emph{et al.} in~\cite{dawson2024swemnics} and GPU accelerated by Pachev  in~\cite{pachev2024automated}, where implicit schemes are considered.   

In Section \ref{sec:model_desc}, we introduce the governing model, i.e., the  SWE and our DG discretization. 
Next, in Sections \ref{sec:GPU_approach} and \ref{sec:experimental_setup}, we discuss the related work and the hardware setup we utilize. Section \ref{sec:implement} details our implementation. In
Section \ref{sec:results}, we discuss multi-node performance results when running the code on a large Hurricane dataset. Finally, in Section ~\ref{sec:conclusion} we draw conclusions and discuss potential future research directions.

\section{Model} \label{sec:model_desc}
\subsection{Shallow water equations}

\begin{figure}[h!]
    \centering
    \begin{tikzpicture}

    \definecolor{waterblue}{RGB}{173,216,230}
    \definecolor{sand}{RGB}{194,178,128}
    
    \draw[black, very thick] (-4,0) -- (4,0);
    
    \draw[brown, thick] plot[smooth, tension=.7] coordinates {(-4,-2.5) (-2,-1.8) (0,-2.3) (2,-2.0) (4,-2.4)};
    \fill[sand] plot[smooth, tension=.7] coordinates {(-4,-3) (-4,-2.5) (-2,-1.8) (0,-2.3) (2,-2.0) (4,-2.4) (4,-3)} -- cycle;

    \draw[blue, thick] plot[smooth, tension=.7] coordinates {(-4,1.0) (-2,0.8) (0,1.2) (2,1.1) (4,1.3)};
    \fill[waterblue] plot[smooth, tension=.7] coordinates {(-4,1.0) (-2,0.8) (0,1.2) (2,1.1) (4,1.3)} -- plot[smooth, tension=.7] coordinates { (4,-2.4) (2,-2.0) (0,-2.3) (-2,-1.8) (-4,-2.5) } -- cycle;
    
    \node[right] at (1, -0.6) {$\quad H = \zeta + h_b$};
    
    \node[right] at (4.5, 0.450) {$\quad \zeta $};
    \node[right] at (4.5, -1.2) {$\quad h_b$};

    \draw[<->] (1.3, 1.1) -- (1.3, -2.1);

    \draw[<->] (4.5, 1.3) -- (4.5, 0);
    \node[above] at (4.5, 1.3) { };

    \draw[<->] (4.5, 0) -- (4.5, -2.4);
    \node[below] at (4.5, -2.4) { };
    
    \draw[black , thick] (-4,0) -- (4,0);

\end{tikzpicture}
    \caption{Definition of shallow water elevations. The horizontal line denotes the geoid, where $\zeta = h_b = 0$.}
    \label{fig:elevation_def}
\end{figure}
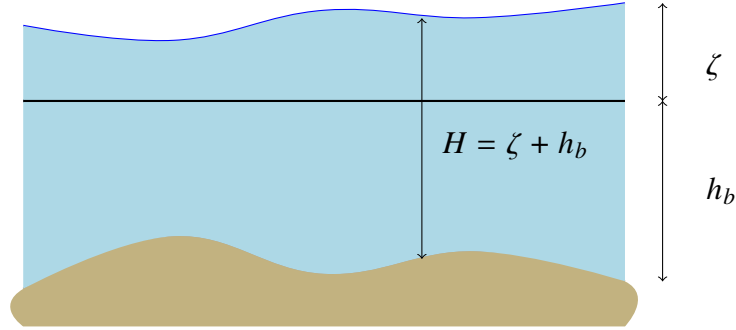
%
%


The governing model in DG-SWEM is the two-dimensional shallow water equations which consists of mass and momentum conservation in each direction. The problem can be visualized as in Figure \ref{fig:elevation_def} and formulated as follows \cite{tan1992shallow}:
\begin{equation} \label{eq:SWE}
\begin{array}{ll}
\text{Find }  (\zeta, \uu) \text{ in } \Omega  \text{ such that:}  \qquad \qquad    \\ \\
\ds \frac{\partial  \zeta}{\partial t} + \nabla \cdot (H{\uu})  = R, \\ \\
\ds\frac{\partial (Hu_x)}{\partial t} + \nabla \cdot \left( Hu_x^2 + \frac{g}{2}(H^2-h_b^2), Hu_xu_y \right) - g\zeta \frac{\partial h_b}{\partial x} + \tau_b Hu_x = F_x,& \\ \\
\ds\frac{\partial (Hu_y)}{\partial t} + \nabla \cdot \left( Hu_xu_y, Hu_y^2 + \frac{g}{2}(H^2-h_b^2) \right) - g\zeta \frac{\partial h_b}{\partial y} + \tau_b Hu_y = F_y,
 \end{array}
\end{equation}
where: 
\begin{description}
    \item[$\zeta$] is the free surface elevation (positive upwards from geoid),
\item[$h_b$] the bathymetry (positive downwards from the geoid), 
\item[$H$] the total water column,  
\item[$\uu = \{ u_x,u_y\}^{\text{T}}$] the depth-averaged velocity field, 
\item[$\tau_b$] the bottom friction factor, 
\item[$R$] the mass source term such as rain,
\item[$F_x,F_y$] the relevant sources which induce flow, including, e.g., Coriolis force, tidal potential forces, wind stresses, and wave radiation stresses, and
\item[$\Omega$]  the computational domain, e.g., the coastal ocean, and its boundary $\Gamma$ is identified by three distinctive sections $\Gamma = \Gamma_{ocean}\cup\Gamma_{land}\cup\Gamma_{river}$. 
\end{description}

\subsection{The Discontinuous Galerkin Formulation}

The particular DG formulation and solution method used in this work is based on  previous works \cite{kubatko2006hp,dawson2011discontinuous}. Hence, we do not include a comprehensive introduction to the details of the formulation but rather a brief overview to make this presentation sufficiently self-contained. This description is an updated version from our recent work \cite{Wichitrnithed2024-jc}.

First, we define the vector of state variables to be solved, $\mathbf{w} = \{\zeta, Hu_x, Hu_y \}^\text{T}$, the source vector $\mathbf{s}$ and the flux matrix $\mathbf{F}(\ww)$ as:
\begin{equation*}
    \mathbf{w} =
                 \begin{bmatrix}
                 \zeta, & uH, & vH
                 \end{bmatrix}^T, \\
\end{equation*}
\begin{equation*}
    \mathbf{s} =
                 \begin{bmatrix}
                   R, & g\zeta\dx{h_b} + F_x,  & g\zeta\dy{h_b} + F_y
                 \end{bmatrix}^T, \\
\end{equation*}
\begin{equation*}
    \mathbf{F}(\ww) =
                 \begin{bmatrix}
                   uH & vH \\
                   Hu^2 + g(H^2-h_b^2) & Huv \\
                   Huv & Hv^2 + g(H^2-h_b^2)
                 \end{bmatrix},
\end{equation*}
we can write these three equations a conservative form
\begin{align} \label{eq:cons_form}
    \dt{\mathbf{w}} + \nabla \cdot \bfm{F}(\ww) = \mathbf{s}, \quad i = 1,2,3.
\end{align}
As is standard in the discretization of \eqref{eq:cons_form} with a  finite element formulation, we partition the domain into (triangular) non-overlapping elements,
$$\Omega = \bigcup_e \Omega_e$$
and the corresponding weak formulation is obtained by multiplying each equation by a test function $\mathbf{v}$ and integrating over each element and subsequently performing an integrating by parts to the term with the spatial derivative:
\begin{align} \label{eq:elem_weak_form}
    \area{\dt{\mathbf{w}}}{\mathbf{v}} = -\area{\nabla \mathbf{v}}{\bfm{F}} - \edge{\hat{\bfm{F}} \cdot \bfm{n}}{\mathbf{v}} + \area{\mathbf{s}}{\mathbf{v}} \quad \text{for each element } e,
\end{align}
where we use standard inner product notation for integration over elements, i.e., $\area{u}{v} = \int_{\Omega_e} u\,v $ and element edges $\edge{u}{v} = \int_{\partial \Omega_e} u\, v$. 
$\hat{\bfm{F}}$ represents the numerical flux, e.g., Local Lax-Friedrichs or Roe flux. The role of the numerical flux is to couple adjacent, discontinuous elements and correctly capture the flow characteristics across the edges; this is necessary for the solution to be stable \cite[Chapter~2]{Hesthaven2008-on}. Thus, aside from calculating $\hat{\bfm{F}}$ (which depends on two neighboring elements), the computation can proceed on each element independently. 

In accordance with the discontinuous Galerkin method, we next approximate the solution $\ww$ in space as $\widetilde{\ww}$ using a set of basis functions. We opt to use the orthogonal Dubiner basis \cite{Dubiner1991-ts} $\phi_{ij}$ as:
\begin{align} \label{dubiner}
    \widetilde{\ww} = \sum_i \sum_j \widetilde{\ww}_{ij} \phi_{ij},
\end{align}
where $\widetilde{\ww}_{ij}$ are the modal degrees of freedom. The indices $i,j$ indicate the order of the polynomial: $\phi_{ij}$ is of order $i+j$. This basis is chosen as it produces a diagonal mass matrix, and higher-order approximations can be obtained by simply adding more terms. Note that while DG-SWEM supports higher-order polynomial orders, this work implements only the linear case, i.e. the first three basis functions are used. In the context of simulating storm surges and similar to ADCIRC, we found that using linear (or even constant) approximation is sufficient. 
Substituting \eqref{dubiner} into the weak formulation (\ref{eq:elem_weak_form}) above, and requiring that it holds for all basis functions $\mathbf{v} \in \{\phi_{00}, \phi_{01}, \phi_{10} \}$ reduces the problem to a system of ODEs in the form:
\begin{align} \label{eq:RKSSP}
    \frac{d}{dt} (\widetilde{\ww}) = L_h (\widetilde{\ww}) \quad \text{on each element } e,
\end{align}
where the terms from~\eqref{eq:elem_weak_form} have been collected into the RHS $L_h$. Numerical integration of each term is performed using Gauss-Lobatto quadrature.

To integrate in time, we use an optimized version of the explicit Strong Stability Preserving Runge-Kutta scheme (SSPRK) based on \cite{Shu1988-eq} to discretize~\eqref{eq:RKSSP} in time and  compute the difference approximation $\ww_h^{n+1}$ of $\widetilde{\ww}$, at the next timestep. This scheme has been specifically optimized with respect to DG spatial discretizations to enhance the CFL stability regions and therefore increase the possible time step size for a given order and stage \cite{Kubatko2014-nc}. This is also combined with an optional slope limiter to suppress oscillations. Finally,
to output the state variables, we convert the modal representation to a nodal representation by averaging the neighboring elements of each node.

In terms of execution, DG-SWEM supports parallel execution on CPUs through domain decomposition with METIS \cite{Karypis1997-kb} using the Message-Passing Interface (MPI). As the scheme is explicit in both space and time, most computations are local to each element, except the calculation of the numerical flux $\hat{\textbf{F}}$ that is shared between two neighboring elements. We use the standard ghost-cell technique to transfer state variables of elements at the boundary of each domain. An evaluation of the parallel efficiency of the code is reported in \cite{Kubatko2009-sn}.

\section{GPU Programming approach} \label{sec:GPU_approach}
As an alternative to purely distributed parallelism, graphics processing units (GPUs) can be viewed as a multithreaded vector processor \cite{Hennessy2018-dw}, where each unit executes the same instructions on multiple data (SIMD), but individual lanes can take different paths. To take advantage of a large number of such processors, the algorithm must exhibit a high degree of data parallelism, where multiple elements can be processed independently at the same time. This is particularly applicable to explicit DG methods due to the inherently local computation of the elements. 

In the past, as GPUs have originally been designed for graphics programming, the available application programming interface (API) such as DirectX and OpenGL imposed limitations of using GPUs for general-purpose computing (GPGPU) \cite{Kirk2010-hd}.
 In 2007, NVIDIA developed CUDA \cite{NVIDIA2024-jw} which provides a more generic API to GPU instructions. This significantly extended the capabilities of NVIDIA GPUs to scientific and engineering codes: the programmer gained control of individual GPU threads, synchronization, and shared memory. Using the NVIDIA HPC compiler stack (a successor of PGI compilers), CUDA is supported in C, C++, and Fortran \cite{NVIDIA2020-vy}; CUDA code can be written in the same source files and compiled with appropriate flags. While the low-level nature of CUDA allows more opportunities for control and optimization by the programmer, it typically results in having to maintain a separate code base from the CPU code. In the past, a framework has been created to generate GPU code from DG-SWEM through CUDA C++ \cite{DuChene2011-ll}. The workflow consisted of scripts that transform Fortran subroutines to C++ and handle variable declaration and allocation. The work noted that other alternatives at the time, such as using the PGI accelerator or CUDA-Fortran, were lacking in features or accessibility.

A more modern approach to GPU programming is to use OpenACC \cite{OpenACC-Standardorg2021-qe} or OpenMP \cite{Openmporg2021-pj} 
 to write \textit{directives} instead of manually managing individual GPU threads. This approach allows the programmer to insert high-level statements that generate GPU instructions and appear as comments in the source code (similar to OpenMP loop constructs, for instance). This can sometimes result in a single code base --- only the compilation flags such as \verb|-acc| need to be turned on to enable the GPU code generation. Of course, the programmer is still required to sufficiently expose the parallelism in the code, which often results in modifications of the CPU code as well. We choose to use OpenACC in this work as it currently has better support on the NVHPC compilers.
 In this approach, we aim to mimimize integration and maintenance efforts, as well as to take advantage of the new NVIDIA Grace-Hopper architecture \cite{Singer-2024} which provides a more efficient CPU-GPU data transfer.

\section{Experimental setup} \label{sec:experimental_setup}
The primary hardware used in this work is the recent Vista Supercomputer \cite{Singer-2024} at the Texas Advanced Computing Center (TACC). Vista uses the ARM-based NVIDIA GH200 Grace Hopper architecture. Each GPU node consists of an NVIDIA H100 GPU connected to an NVIDIA Grace CPU directly through high-bandwidth NVLink-C2C interconnect. The Grace CPU contains 72 Arm Neoverse V2 cores and 120 GiB of LPDDR5X memory, while the Hopper GPU contains 96 GB of HBM3 memory. 
Each CPU node in the CPU-only subsystem contains 144 Grace cores on two sockets with a total memory of 237 GB LPDDR5. Different nodes are connected in a fat-tree topology, with GG nodes through an InfiniBand 200 GB/s and GH nodes an InfiniBand 400 GB/s. Specifications for these nodes are listed in Table \ref{tab:gh} and \ref{tab:gg}.

The NVLink interconnect provides a faster data transfer between CPU and GPU. Previous generations of GPUs connect to CPUs through PCIe and trigger data migration on page faults. GH200 uses a hardware-based approach to achieve the same functionality by sharing the same page table and virtual address space which results in less page faults. In other words, CPU and GPU memories are now treated as NUMA nodes \cite{Fusco2024-uh}.

\begin{table}[!ht]
    \centering
    \begin{tabular}{l l}
    \hline
        CPU & NVIDIA Grace, 72 cores @ 3.1 GHz \\
        CPU memory & 120 GB LPDDR5X \\
        \hline
         GPU & NVIDIA H100 \\
         No. of SMs & 132 \\
         FP64 & 34 TFLOPs/s \\
         FP32 & 67 TFLOPs/s \\
         GPU memory & 96 GB HBM3 \\
         GPU memory bandwidth & 4 TB/s \\
         NVLink-C2C CPU-GPU bandwith & 900 GB/s \\
         \hline
    \end{tabular}
    \caption{A Grace Hopper (GH) GPU node}
    \label{tab:gh}
\end{table}

\begin{table}[h]
    \centering
    \begin{tabular}{l l}
    \hline
        CPU & NVIDIA Grace, 144 cores @ 3.4 GHz \\
        FP64 & 7 TFLOPs/s \\
        CPU memory & 237 GB LPDDR5 \\
         CPU memory bandwidth & 850 GiB/s \\
         \hline
    \end{tabular}
    \caption{A Grace Grace (GG) CPU node}
    \label{tab:gg}
\end{table}
The original CPU code was compiled using the Intel oneAPI compiler stack. This work uses the NVIDIA SDK compiler stack 24.7 (\verb|nvfortran|, \verb|nvc, nvc++|) for both CPU and GPU versions.  Compilation flags for the CPU and GPU codes are listed in Table \ref{tab:flags} .

\begin{table}[h]
    \centering
    \begin{tabular}{l  l}
    \hline
    Version & Flags \\
    \hline
       GPU  &  mpif90 -r4 -O3 -cuda -acc -Mextend -Mpreprocess -gpu=unified \\
        CPU &  mpif90 -r4 -O3 -Mextend -Mpreprocess \\
        \hline
    \end{tabular}
    \caption{Compiler flags used for the single-precision option.}
    \label{tab:flags}
\end{table}

\section{Implementation} \label{sec:implement}

\subsection{Unified Memory}
A significant part of porting code to GPUs is managing data movement. As a CPU (host) and a GPU (device) have separate physical memories, data must be transferred back and forth to perform the respective calculations. Traditionally, even if some arrays are only used in device calculations, they still need to be initialized or read from files through the host, and we would have to declare device copies of those arrays. For example, in CUDA Fortran, an existing array in the code

\verb|real, allocatable :: A(:)|

\noindent
will need an additional copy on the device

\verb|real, allocatable, device :: A_d(:)|

\noindent
and copied to and from the device as necessary:

\verb|A_d = A|

\verb|! Use A_d on the GPU|

\verb|A = A_d|

\noindent
To avoid having to do this for each array in the code, we exploit
Unified Memory which was introduced in CUDA 6.0. This allows GPU kernels to access data on the host without needing to explicitly copy the data to device. Data migration is automatically performed once a page fault is triggered on the device. In the past, this requires a special allocation, e.g. \verb|cudaMallocManaged| in CUDA C.
For newer systems that support Heterogenous Memory Management (HMM), all device data declarations can be completely eliminated using the compilation flag \verb|-gpu=unified|. This significantly simplifies programming effort and enhances maintainability. In both approaches, a synchronization statement like \verb|!$acc wait| is needed for correctness before the CPU can access the computed data, since GPU kernels immediately return control to the host. 

In Table \ref{tab:arrays} we summarize the important data arrays, some of which are referred to in the kernel descriptions in the next section.

\begin{table}[h]
\footnotesize
    \centering
    \begin{tabular}{l l}
    \hline\hline
       Array  & Meaning   \\
       \hline
        ZE[k,e,irk] & $k^{th}$ mode of elevation solution at element $e$ on RK stage irk  \\
        QX[k,e,irk] & $k^{th}$ mode of x-velocity solution at element $e$ on RK stage irk  \\
        QY[k,e,irk] & $k^{th}$ mode of y-velocity solution at element $e$ on RK stage irk \\
        RHS\_ZE[k,e,irk] & Contribution to the $k^{th}$ mode of $d\zeta/dt$ at element $e$ on RK stage irk  \\
        RHS\_QX[k,e,irk] & Contribution to the $k^{th}$ mode of $dQ_x/dt$ at element $e$ on RK stage irk  \\
        RHS\_QY[k,e,irk] & Contribution to the $k^{th}$ mode of $dQ_y/dt$ at element $e$ on RK stage irk  \\
        PHI\_AREA[k,i] & $k^{th}$ basis evaluated at area quadrature point i  \\
        PHI\_EDGE[k,i] & $k^{th}$ basis evaluated at edge quadrature point i  \\
        XFAC[k,i,e] & Weight for the $k^{th}$ basis of element $e$ in x-direction at area quadrature point i  \\
        YFAC[k,i,e] & Weight for the $k^{th}$ basis of element $e$ in x-direction at area quadrature point i  \\
        SRFAC[k,i,e] & Weight for the $k^{th}$ basis of the source term on element $e$ at area quadrature point i \\
        EDGEFAC[k,i,l] & Weight for the $k^{th}$ basis of the source term on element $e$ at area quadrature point i \\
        EDGE\_ze[k,l,e] & Numerical flux contribution to the $k^{th}$ basis of $\hat\zeta$ at local edge $l$ of element $e$  \\
        EDGE\_qx[k,l,e] & Numerical flux contribution to the $k^{th}$ basis of $\hat{Q_x}$ at local edge $l$ of element $e$  \\
        EDGE\_qy[k,l,e] & Numerical flux contribution to the $k^{th}$ basis of $\hat{Q_y}$ at local edge $l$ of element $e$  \\

        \hline\hline
    \end{tabular}
    \caption{Primary data arrays in the code.}
    \label{tab:arrays}
\end{table}

\subsection{GPU kernels}
This section outlines the main subroutines and how their parallelization. For the integration kernels, we follow the one-thread-per-element and one-thread-per-edge approach as in \cite{Fuhry2013-xg}.

\subsubsection{Area integration}
The area integration routine (Algorithm \ref{alg:rhs}) is naturally data-parallel, as computations are done locally to each element $\Omega_e$. At a given Runge-Kutta stage \verb|irk|, it computes the terms
\begin{align}
    \area{\nabla v}{F} + \area{s}{v}
\end{align}
from \eqref{eq:elem_weak_form} and accumulates them into the RHS vectors. 
Recall from Eq \ref{eq:cons_form} that the flux terms (explicitly written in terms of the state variables) are
\begin{align}
    F_x, F_y &= [Q_x,\quad Q_y], \\
    G_x, G_y &= \left[\frac{Q_x^2}{(\zeta+h)} + \frac{1}{2}g((\zeta+h)^2-h^2), \quad \frac{Q_xQ_y}{(\zeta+h)} \right], \\
    H_x, H_y &= \left[\frac{Q_xQ_y}{(\zeta+h)}, \quad \frac{Q_y^2}{(\zeta+h)}+\frac{1}{2}g((\zeta+h)^2-h^2) \right],
\end{align}
and these are represented by the procedures \textsc{Continuity\_Flux}, \textsc{X\_Flux}, and \textsc{Y\_Flux} in the code. The source terms $S_R, S_x, S_y$ are calculated in the same way but retrieve additional variables such as current rain intensity and wind speed.

There are technically three levels of loop parallelism: looping over the elements, looping over the quadrature points, and looping over the degrees of freedom. Here, the element-level loop is accelerated through the gang-vector parallelism over the number of elements $N_e$ (line 2), and the inner loops over quadrature points and degrees of freedom are enforced to be serial using \verb|!$acc seq|. This needs to be explicitly inserted, as by default the inner loops will also be parallelized, and in this case the trip counts is too short to be profitable. We also add the \verb|private(...)| clause to keep any local arrays private to each thread.
Importantly, the \verb|async| clause is added to eliminate kernel launch overhead, and the \verb|default(present)| the overhead of checking if data is already on the device. This same structure is applied for the edge integration kernel. We have observed that using these clauses speeds up the program by almost 1.5$\times$.

\begin{algorithm}[ht!]
\footnotesize
\caption{Area integration (AREA)}
\label{alg:rhs}
    \begin{algorithmic}[1]
    \Procedure{Rhs\_Dg\_Hydro}{irk}

   \LComment{!\$acc parallel loop gang vector private(...) async(1) default(present)}
    \For{L = 1 ... $N_e$}
\LComment{!\$acc loop seq}
    \For{i = 1 ... quadrature points}
    \State $\zeta \gets 0$
        \State $Q_x \gets 0$
    \State $Q_y \gets 0$
\LComment{!\$acc loop seq}
    \For{k = 1 ... degrees of freedom}
    \State $\Phi \gets $ PHI\_AREA[k,i]
    \State $\zeta \gets \zeta$ + ZE[k,L,IRK] $\times \ \Phi$
    \State $Q_x \gets Q_x$ + QX[k,L,IRK] $\times \ \Phi$
    \State $Q_y \gets Q_y$ + QY[k,L,IRK] $\times \ \Phi$
    \EndFor
    \State

        \State $F_x, F_y \gets $ \Call{Continuity\_Flux}{$\zeta$, $Q_x$, $Q_y$}
        \State $G_x, G_y \gets $ \Call{X\_Flux}{$\zeta$, $Q_x$, $Q_y$}
        \State $H_x, H_y \gets $ \Call{Y\_Flux}{$\zeta$, $Q_x$, $Q_y$}

\State
        \State $S_R \gets $ \Call{Continuity\_Source}{} \Comment{Rain}
        \State $S_x \gets $ \Call{X\_Source}{$\zeta$, $Q_x$, $Q_y$} \Comment{Wind, bottom friction, etc.}
        \State $S_y \gets $ \Call{Y\_Source}{$\zeta$, $Q_x$, $Q_y$} \Comment{Wind, bottom friction, etc.}

        \State
\LComment{!\$acc loop seq}
    \For{k = 1 ... degrees of freedom}
    \LComment{Integration weights}
    \State $W_x \gets$ XFAC[k,i,L]
    \State $W_y \gets$ YFAC[k,i,L]
    \State $W_s \gets$ SRFAC[k,i,L]

\State
    \State RHS\_ze[k,id,irk] $\gets$ RHS\_ze[k,L,irk] + $W_x \cdot F_x + W_y \cdot F_y + W_s \cdot S_R$
    \State RHS\_qx[k,id,irk] $\gets$ RHS\_qx[k,L,irk] + $W_x \cdot G_x + W_y \cdot G_y + W_s \cdot S_x$
    \State RHS\_qy[k,id,irk] $\gets$ RHS\_qx[k,L,irk] + $W_x \cdot H_x + W_y \cdot H_y + W_s \cdot S_y$
    \EndFor
    \EndFor

    \EndFor
    \EndProcedure
    \end{algorithmic}
\end{algorithm}

\subsubsection{Edge integration}
Edge integration routines compute the line integrals
\begin{align}
    \edge{\hat{\bfm{F}}_i \cdot \bfm{n}}{\mathbf{v}}, \quad i=1,2,3,
\end{align}
and accumulate them to the RHS. Similar to the volume integration kernel, a natural parallelism is in the number of edges $N_d$. There are multiple ways to add contributions from inter-element fluxes. One is to let each edge compute the numerical flux between the two elements, and then atomically add back the contributions to each. Atomics are necessary since two edges can share the same element. 

Another approach is to first store the flux contributions to each element separately for each of its edges, i.e., we create an additional array \verb|elem_edge(E, i)| which stores the contribution at edges $i = 1,2,3$ of element $E$. Subsequently, in each element we then sum the three contributions locally without needing atomics. We use this approach in our work as it leads to reproducible, deterministic results due to the fixed order of summation. The pseudocode for these two steps are shown in Algorithm \ref{alg:internal} and Algorithm \ref{alg:flux}. Note that we have abstracted away the wetting/drying branching for brevity.

In Algorithm \ref{alg:internal}, each GPU thread is mapped to an interior edge, again through gang-vector parallelism (line 2). In line 4-9, the edge number is used to obtain the IDs of the two neighboring elements as well as their local edge numbers (1-3), with the interior side denoted with $+$ and exterior $-$. Line integration on the edge is done by looping through the edge quadrature points. First, at each point we compute the state variables (Line 14-23), use them to compute the numerical flux. Then, we add this flux contributions to both sides of the edge, indexed using the local edge number of each element (line 35-41). The subsequent elemental flux summation is shown in Algorithm \ref{alg:flux}, where each thread is mapped to an element and sums the contributions from its 3 edges. Note that for brevity, we have opted for the atomic approach for other less common boundary integration routines.

\begin{algorithm}[ht!]
\footnotesize
\caption{Interior edge integration (EDGE)}
\label{alg:internal}
    \begin{algorithmic}[1]
    \Procedure{Internal\_Edge\_Hydro}{irk}

\LComment{!\$acc parallel loop gang vector private(...) async(1) default(present)}
\For{D = 1 ... $N_d$}

    \State $L \gets $ NIEDN[id] \Comment{Global edge number}
    \State $e^+ \gets $ NEDEL[1,L] \Comment{ID of interior element}
    \State $e^- \gets $ NEDEL[2,L] \Comment{ID of exterior element}
\State
\State $l^+ \gets $ LED[1,L] \Comment{Local edge number of interior element}
    \State $l^- \gets $ LED[2,L] \Comment{Local edge number of exterior element}
\LComment{!\$acc loop seq}
    \For{i = 1 ... edge quadrature points}
    \State $\zeta^{\pm}, Q_x^{\pm}, Q_y^{\pm} \gets 0$
       
\LComment{!\$acc loop seq}
    \For{k = 1 ... degrees of freedom}
    \State $\Phi \gets $ PHI\_EDGE[k,$i^+$]
    \State $\zeta^+ \gets \zeta^+$ + ZE[k,$e^+$,IRK] $\times \ \Phi$
    \State $Q_x^+ \gets Q_x^+$ + QX[k,$e^+$,IRK] $\times \ \Phi$
    \State $Q_y^+ \gets Q_y^+$ + QY[k,$e^+$,IRK] $\times \ \Phi$
\State
     \State $\Phi \gets $ PHI\_EDGE[k,$i^-$]
    \State $\zeta^- \gets \zeta^-$ + ZE[k,$e^-$,IRK] $\times \ \Phi$
    \State $Q_x^- \gets Q_x^-$ + QX[k,$e^-$,IRK] $\times \ \Phi$
    \State $Q_y^- \gets Q_y^-$ + QY[k,$e^-$,IRK] $\times \ \Phi$
    \EndFor
    \State
        \State $\hat{{F}}, \hat{{G}}, \hat{{H}} \gets $ \Call{Numerical\_Flux}{$\zeta^+,\zeta^-,Q_x^+,Q_x^-,Q_y^+,Q_y^-$}

        \State
\LComment{!\$acc loop seq}
    \For{k = 1 ... degrees of freedom}
    \LComment{Integration weights}
    \State $W^+ \gets$ EDGEFAC[k,i1]
    \State $W^- \gets$ EDGEFAC[k,i2]

\State
\LComment{Store flux contributions at the local edge number of each element}
    \State EDGE\_ze[$e^+$,$l^+$,k] $\gets$ EDGE\_ze[$e^+$,$l^+$,k] + $W^+ \cdot \hat{{F}}$
  \State EDGE\_qx[$e^+$,$l^+$,k] $\gets$ EDGE\_qx[$e^+$,$l^+$,k] + $W^+ \cdot \hat{{G}}$
    \State EDGE\_qy[$e^+$,$l^+$,k] $\gets$ EDGE\_qy[$e^+$,$l^+$,k] + $W^+ \cdot \hat{{H}}$

    \State
    \State EDGE\_ze[$e^-$,$l^-$,k] $\gets$ EDGE\_ze[$e^-$,$l^-$,k] + $W^- \cdot \hat{{F}}$
  \State EDGE\_qx[$e^-$,$l^-$,k] $\gets$ EDGE\_qx[$e^-$,$l^-$,k] + $W^- \cdot \hat{{G}}$
    \State EDGE\_qy[$e^-$,$l^-$,k] $\gets$ EDGE\_qy[$e^-$,$l^-$,k] + $W^- \cdot \hat{{H}}$
    \EndFor
    \EndFor

    \EndFor
    \EndProcedure
        \end{algorithmic}
\end{algorithm}

\begin{algorithm}
\footnotesize
\caption{Flux gathering kernel (GATHER)}
\label{alg:flux}
    \begin{algorithmic}[1]
        \Procedure{Flux\_Gather}{irk}

    \LComment{!\$acc parallel loop gang vector async(1) default(present)}
    \For{L = 1 ... $N_e$}
    \LComment{!\$acc loop seq}
    \For{k = 1 ... degrees of freedom}
    \For{i = 1 ... 3}
    \State RHS\_ze[k,L,irk] $\gets$ RHS\_ze[k,L,irk] + EDGE\_ze[L,i,k]
        \State RHS\_qx[k,L,irk] $\gets$ RHS\_qx[k,L,irk] + EDGE\_qx[L,i,k]
    \State RHS\_qy[k,L,irk] $\gets$ RHS\_qy[k,L,irk] + EDGE\_qy[L,i,k]

    \EndFor
    \EndFor
    \EndFor
    \EndProcedure
        \end{algorithmic}
\end{algorithm}

\subsubsection{Other kernels}
Other kernels like slope limiting (SLOPELIM) and wet/dry calculations (WETDRY) are similarly parallelized at the element-level loop, with small inner loops serialized. Wind forcing and pressure calculations (WIND) are done in a series of loops, each operating on the mesh nodes independently. We simply parallelize each loop with \texttt{!\$acc parallel loop}. 

\subsubsection{Data transfers between host and device}
At every fixed number of time steps specified by the user, the program writes the state variables to ASCII files. Thus data will have to be copied to the CPU. While it is possible to completely rely on HMM to initiate the transfers, they can incur large overhead as there are enough access counts to trigger page migration. To avoid this, we use \texttt{cudaMemPrefetchAsync} to transfer data at once, as shown in Listing \ref{lst:io}.

\begin{lstlisting}[caption={Data prefetching example for writing out the water elevation array ETA2. The array is moved back to the GPU once the writing is done.}, language=Fortran,label={lst:io}]
!@cuf use cudafor
!$acc wait
!@cuf istat = cudaMemPrefetchAsync(eta2, np, cpuDeviceID, 1) 
do i = 1, np
    write(63, *) i, eta2(i)
enddo
!@cuf istat = cudaMemPrefetchAsync(eta2, np, 0, 0)     
    
\end{lstlisting}

\subsubsection{Memory usage of automatic arrays}
We sometimes observe runtime memory error when parallelizing loop with many automatic (private) arrays per thread. From our observation, this seems to be due to the allocation on these arrays on the GPU stack, a different behavior from using the CUDA API. We can avoid this by limiting the number of gangs to a fixed number, e.g.

\texttt{!\$acc parallel loop gang vector num\_gangs(1024) private(<long list of arrays>)}.

\noindent
as opposed to allowing an automatic scheduling of gangs.

\subsection{Multiple GPUs}
To exchange data among different domains when using multiple processors, DG-SWEM uses MPI persistent communication. At the start of the simulation, calls to \verb|MPI_SEND_INIT()| and \verb|MPI_RECV_INIT()| are made to specify the location of data to exchange to each neighbor. Then, as shown in Listing \ref{code:mpi}, whenever we need to exchange data, we call \verb|MPI_STARTALL()| to initiate the communication in a non-blocking manner. At the end, we wait until all the messages have arrived. The packing of the elements to send is done through the subroutine \verb|packElems()|, and the unpacking similarly through \verb|unpackElems()|.

The existing MPI routines need no further modification in the GPU code, as they accept unified memory addresses. However, we observe data being transferred to host and back to device during MPI communication. This appears to be a limitation of using unified memory with CUDA-aware MPI which prevents direct GPU-to-GPU data transfer \cite{Chenbr_2025}. The packing and unpacking have been parallelized to take advantage of the higher memory bandwidth of the GPU, as shown in Listing \ref{code:pack}.

\begin{lstlisting}[language=Fortran,
caption={Sending and receiving elements from neighbors. nelemsend(j) is the number of elements to send to neighbor j, send\_indices(:,j) contains the location of those elements; vec is the state variable to send, and sendbuf is the contiguous buffer that will contain the packed data. Receiving uses a similar notation.},
label=code:mpi]
!$acc wait
do j = 1,neighbors_to_send
    call packElems(nelemsend(j), send_indices(:,j), vec, sendbuf(:,j))
enddo
!$acc wait
call MPI_STARTALL(...)

do while (...)
    call MPI_Waitsome(...)
    call unpackElems(nelemrecv(j), recv_indices(:,j), recvbuf(:,j), vec)
enddo
!$acc wait

\end{lstlisting}

\begin{lstlisting}[language=Fortran,
caption={Subroutines for packing and unpacking of ghost elements, where each element consists of DOF points.}, label=code:pack]
subroutine packElems(nsend, send_indices, vec, sendbuf)
!$acc parallel loop default(present) async(1)
do i = 1,nsend
    j = send_indices(i)
    packed((i-1)*DOF+1:i*DOF) = vec(1:DOF,j)
enddo
end subroutine

subroutine unpackElems(nrecv, recv_indices, recvbuf, vec)
!$acc parallel loop default(present) async(1)
do i = 1,nrecv
    j = recv_indices(i)
    vec(1:DOF,j) = recvbuf((i-1)*DOF+1:i*DOF)
enddo
end subroutine
\end{lstlisting}

\section{Results} \label{sec:results}
\subsection{Test Case}
\begin{figure}
    \centering
    \includegraphics[width=0.7\linewidth]{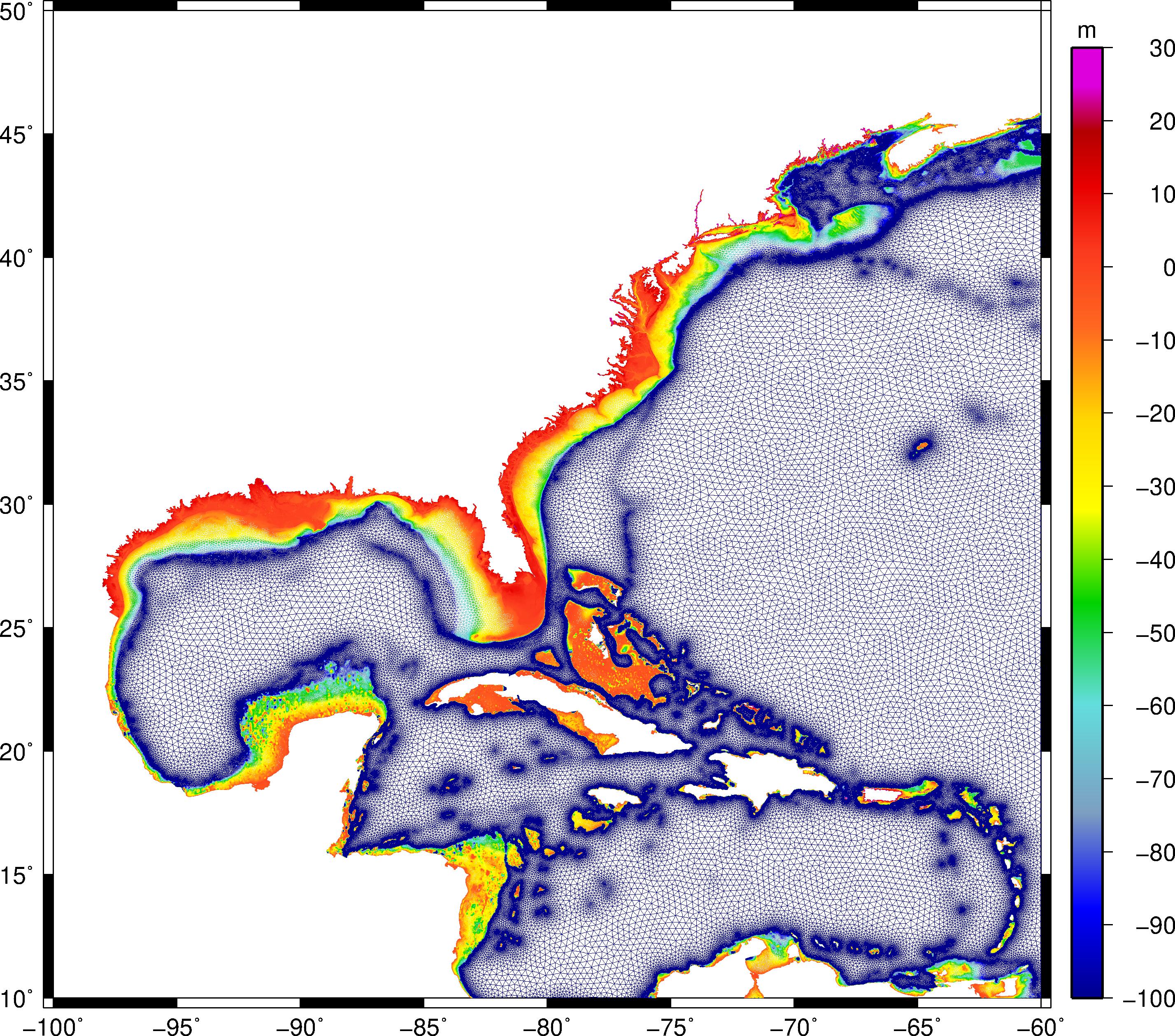}
    \caption{TX2008 mesh \protect\cite{hope2013hindcast} used in the Harvey scenario. Color represents bathymetry, negative downwards.}
    \label{fig:2008}
\end{figure}

\begin{figure}
    \centering
    \includegraphics[width=0.9\linewidth]{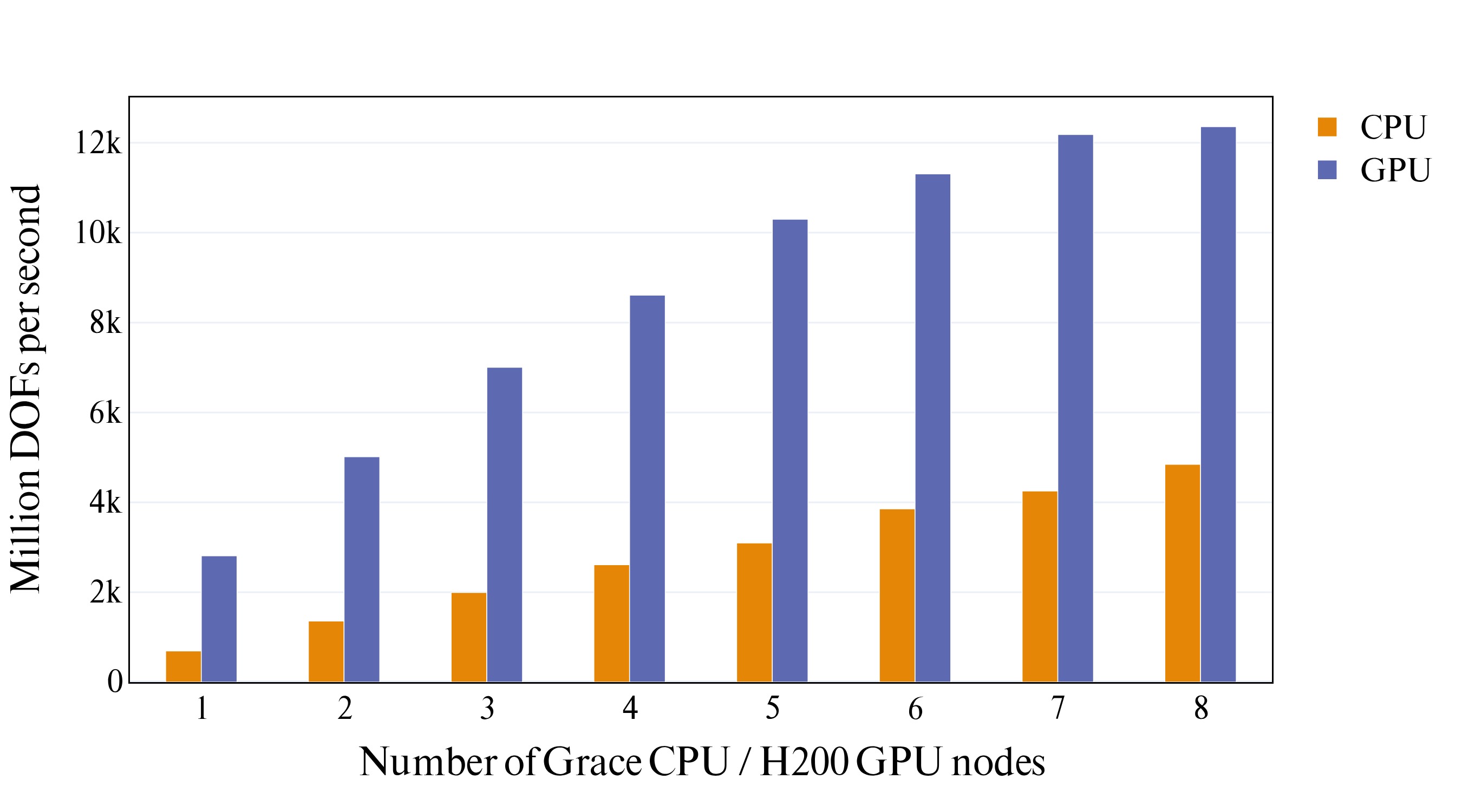}
    \caption{Number of degrees of freedom computer per second for different numbers of compute nodes. Because we use $p=1$ elements, each element has 9 DOFs in total: 3 modal coefficients for each state variable $\zeta,u_x,u_y$.}
    \label{fig:strong}
\end{figure}

\begin{figure}[h!]
    \centering
    \includegraphics[width=0.9\linewidth]{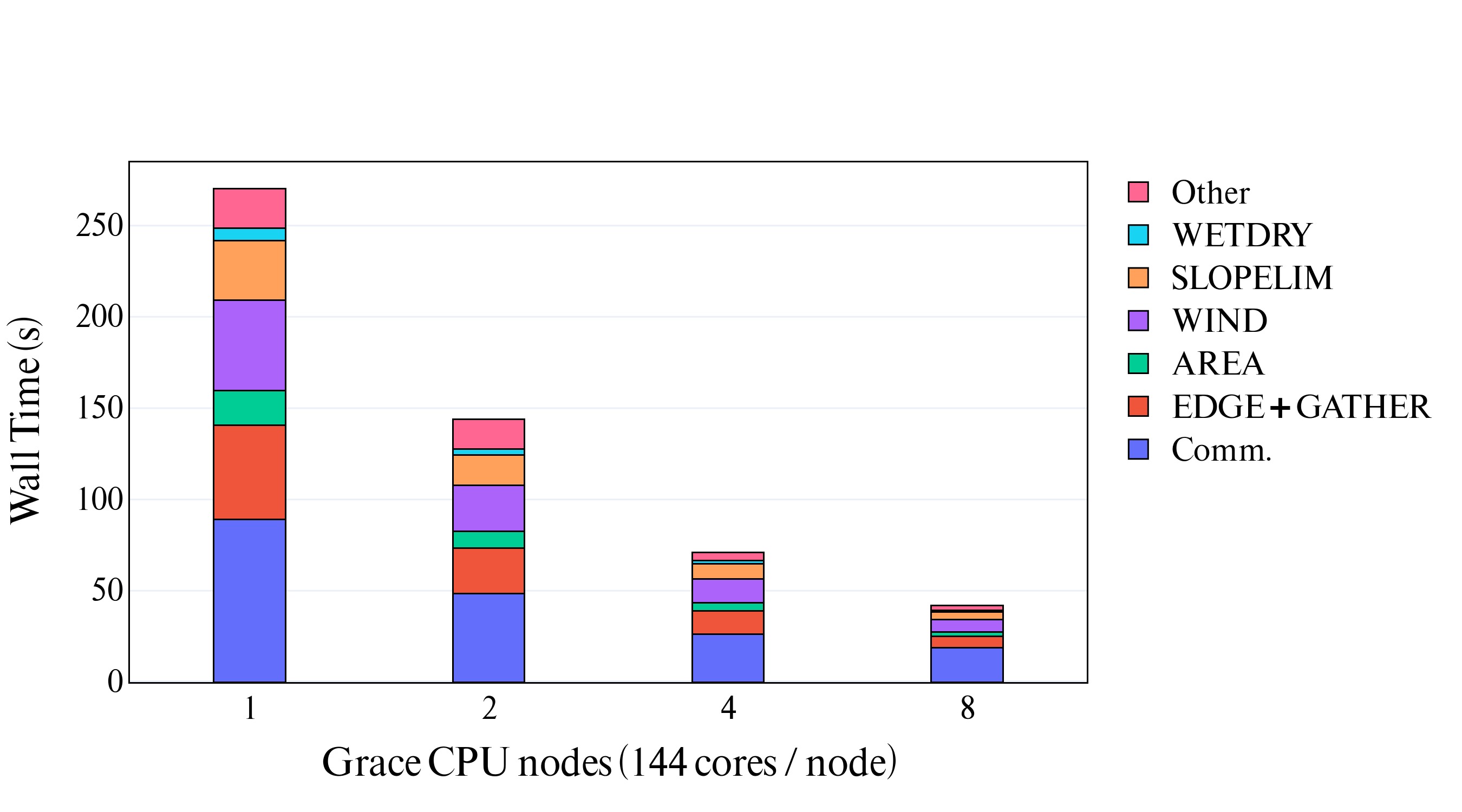}\\
    \includegraphics[width=0.9\linewidth]{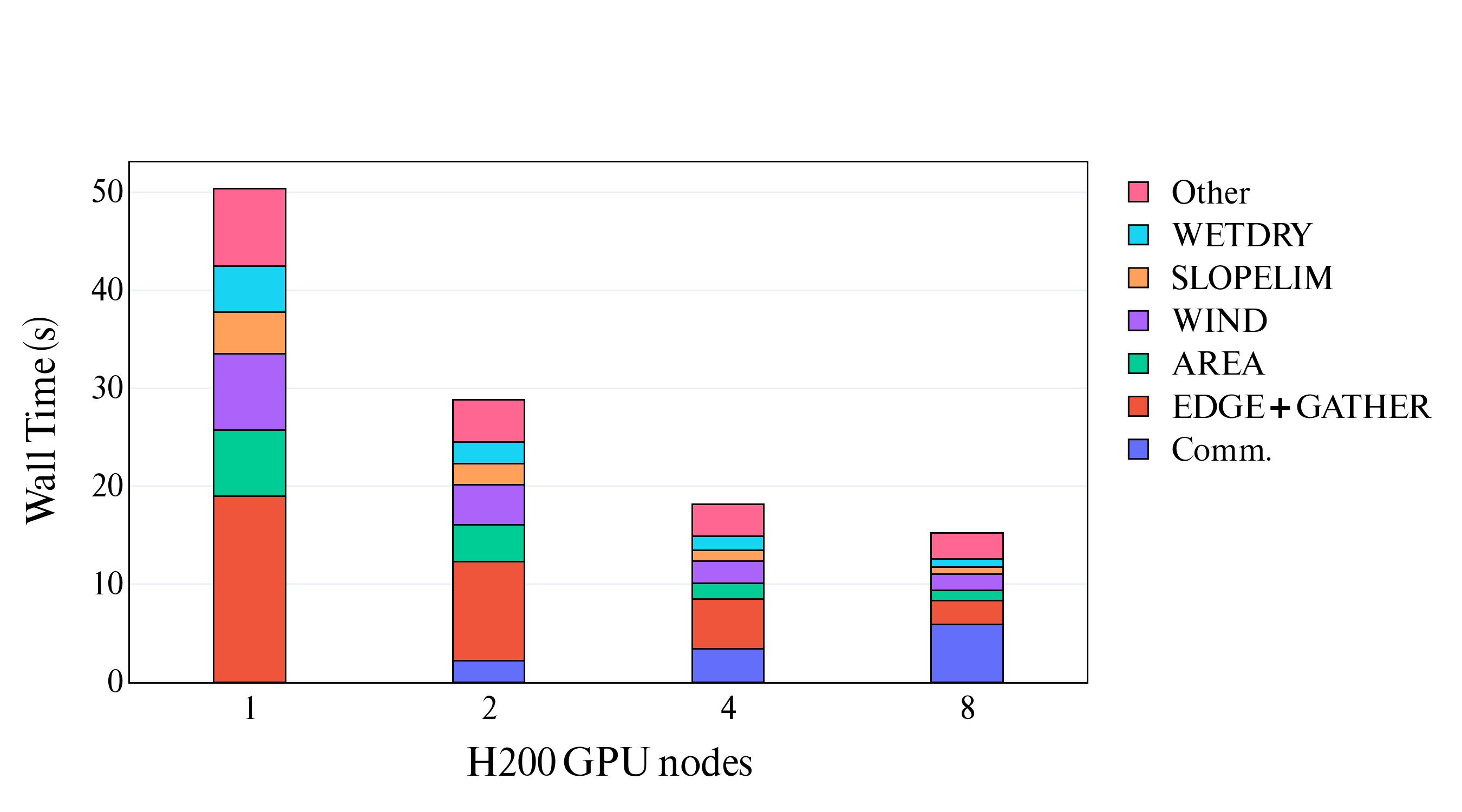}\\

    \caption{Breakdown of time spent (averaged over CPU cores/GPUs) in each computational kernel when running for 2,880 time steps. "Comm." refers to MPI communication routines, including packing and unpacking of data.}
    \label{fig:breakdown}
\end{figure}

To measure performance improvements, we run the solver on a dataset for  Hurricane Harvey (2017) which is representative of a typical simulation scenario.
The mesh is shown in Figure \ref{fig:2008} which contains 6,675,517 elements and 3,352,598 nodes with a maximum resolution of 20 meters.   We impose a tidal (periodic) elevation boundary on the right side and use Harvey Best Track data from the National Hurricane Center HURDAT2 database. The timestep size is set to be $\Delta t=0.3$s, and computation is done with single-precision. 
Figure \ref{fig:strong} shows the performance comparison between the GPU implementation and the CPU implementation from 1 to 8 nodes, where we run the simulation for 28,800 time steps. As described in Section \ref{sec:experimental_setup}, each GPU node consists of a single NVIDIA H200, while each CPU node contains 144 NVIDIA Grace cores. Here, "performance" is measured by the rate of work done, defined to be the number of degrees of freedom computed per second. In a node-by-node comparison, we observe a speedup of GPU code over CPU of around $4\times$ to $2.55\times$.
Figure \ref{fig:breakdown} further shows the breakdown of time spent in each kernel. In this case, we only consider the total time spent in the timestepping loop (i.e. exclude I/O or startup time) and run the simulation for 2,880 time steps.
Here, we see that parallel communication time in the GPU case increases with more GPUs to around 6 s at 8 nodes, resulting in a scaling plateau. For the same amount of nodes in the CPU case case, the communication time appears to stabilize to around 18 s. 

\subsection{Performance Evaluation}
To further evaluate performance of the time-consuming kernels, we first calculate the theoretical arithmetic intensity $\alpha$ at the DRAM level. 
This is the ratio of floating point operations (FLOPs) to total memory traffic in bytes from/to DRAM:
\begin{align}
    \alpha = \frac{\text{Floating point operations}}{\text{Bytes read from DRAM }  + \text{Bytes written to DRAM } }.
\end{align}
We count each arithmetic operation like addition and division counts as one FLOP, including special operations like square root and power. 
In estimating $\alpha$ from the source code (by manually counting), we also assume that everything is loaded once from memory for simplicity, which is generally not valid for large enough datasets and when accesses are not fully coalesced. Thus we expect the estimated values to be higher than the true values.
The estimated arithmetic intensities for the integration kernels are shown in Table \ref{tab:ai}, along with their experimental values obtained from the NVIDIA Nsight Compute profiler. We focus on these two kernels as they are representative of the DG method. 
\begin{table}[h!]
    \centering
    \begin{tabular}{c c c c | c}
    \hline
     & & Estimated & &  Experimental \\
    \hline
        \textbf{Kernel} & \textbf{GFLOPs} & \textbf{GBytes} & $\alpha$ &  $\alpha$  \\
         \hline
    EDGE & 7.95 & 1.6 & 4.97 &  2.7 \\
    AREA & 4.09 & 2.2 & 1.85 &  1.81 \\
    GATHER & 0.24 & 1.92 & 0.125 &  0.09 \\
    \hline
    \end{tabular}
    \caption{Estimated and experimental DRAM arithmetic intensities $\alpha$ for several kernels. Values for GFLOPs and Gbytes represent the total count for all elements/nodes/edges for the Harvey test case.}
    \label{tab:ai}
\end{table}
It can be seen that the arithmetic intensity of AREA is very close to the estimated value. This is expected, as most data are not shared between different elements, and are loaded in a (mostly) coalesced manner. A similar reasoning applies for GATHER. On the other hand, EDGE has a markedly lower intensity than estimated, a little over half. This can be attributed to the different memory access patterns. While area integration loops over elemental arrays sequentially, edge integration requires each edge to fetch data from neighboring elements which may not lie close together in memory. This poor spatial locality is a well-known issue in using unstructured meshes \cite{Hadade2020-fo} and in our case results in uncoalesced accesses and therefore more data traffic. 

We further evaluate the performance using the empirical Roofline which prescribes a performance bound through the relation:
\begin{align*}\label{eq:roofline}
    \text{Attainable performance (FLOPs/s)} = \min{ (\text{Peak FLOPs/s}, \;
    \alpha \times \text{Peak DRAM bandwidth})},
\end{align*}
which makes some assumptions, e.g., latencies are ignored and computation perfectly overlaps with data transfer \cite{Hager2010-qa}. In kernels with low $\alpha$ (like most in our code), performance will, at best, reach only a fraction of the peak FLOP/s of the GPU. 

\begin{figure}[h!]
    \centering
    \includegraphics[width=0.9\linewidth]{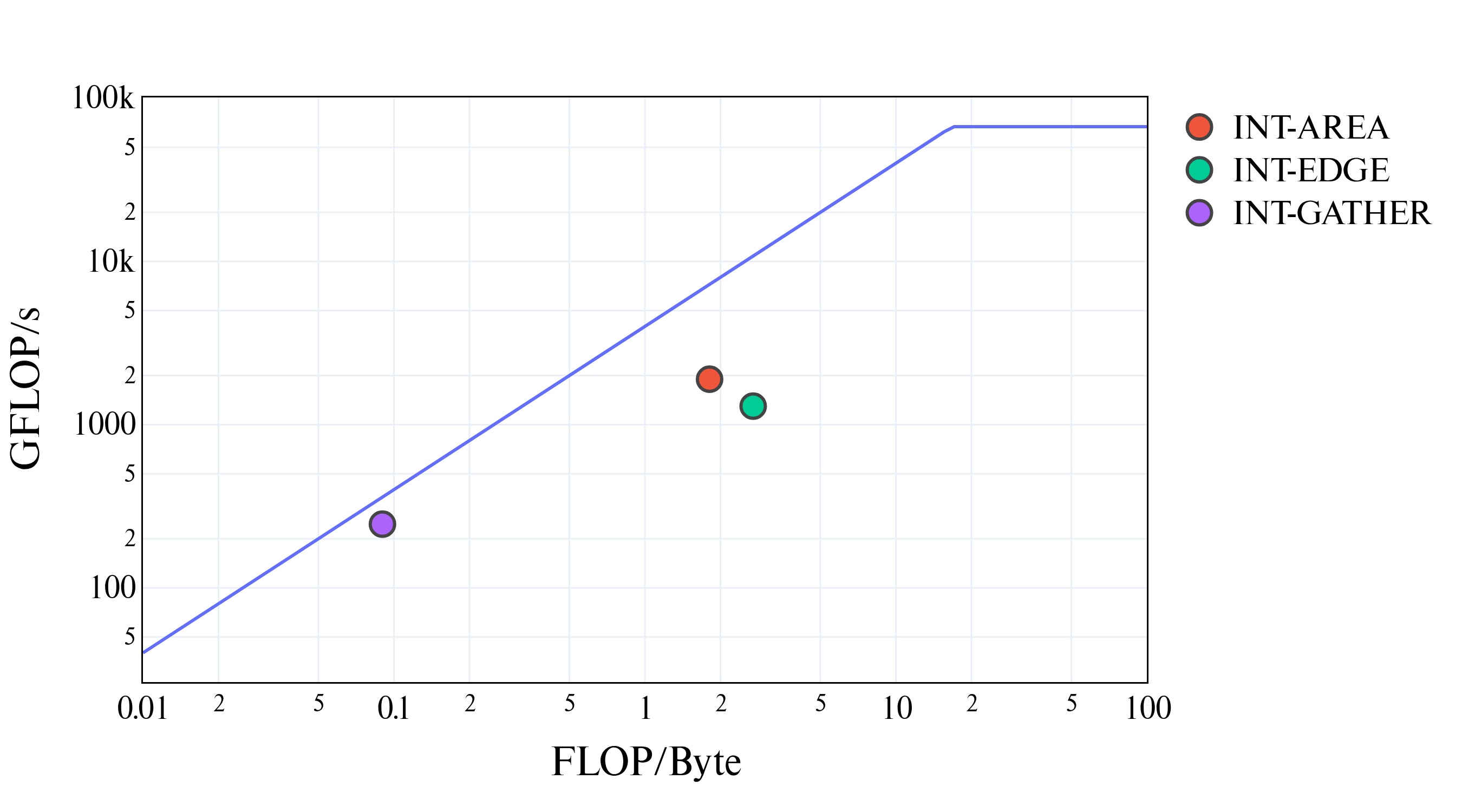}
    \caption{Empirical DRAM single-precision roofline for the integration kernels.}
    \label{fig:roofline}
\end{figure}

To plot the empirical roofline, we use the procedure described in \cite{Yang2020-zg}; this is shown in
Figure \ref{fig:roofline}. The AREA, EDGE, and GATHER kernels achieve compute throughputs of 1.9 TFLOPs/s, 1.3 TFLOPs/s, and 0.24 TFLOPs/s, respectively. Other relevant performance characteristics from Nsight Compute are shown in Table \ref{tab:nsight}. A few observations can be made. First, the INT instruction count shows that a large portion of the computation are in fact array index calculations. While this is largely unavoidable, they do lower the measured FLOP rate, as we cannot assume they overlap with FP32 instructions. Second, the achieved occupancy for the first two kernels are quite low due to the high register usage, and this can also limit performance. While it is possible to tune register usage through the \texttt{-maxregcount} compiler flag, we found that the penalty from spilling largely offsets the speedup from higher occupancy. We also note that using linear ($p = 1$) elements is typically insufficient to reach close to peak performance of   the GPU \cite{Fuhry2013-xg,Klockner2009-is,Abdi2019-rs}.

\begin{table}[h!]
    \centering
    \begin{tabular}{c c c c c}
    \hline
        Kernel & Registers per thread & Occupancy (\%) & FP32 Inst. & INT Instr. \\
        \hline
        EDGE & 249 & 9.4 & 197 $\times 10^6$ & $118 \times 10^6$\\
        AREA & 163 & 14.9 & $110 \times 10^6$ &  $78 \times 10^6$\\
        GATHER & 72 & 40.6 & $5.6 \times 10^6$ &  $110 \times 10^6$\\

         \hline
    \end{tabular}
    \caption{Occupancy and instruction breakdown of the integration kernels. Note that these are \textit{warp instructions}, i.e. each one operates on 32 scalars.}
    \label{tab:nsight}
\end{table}

\subsection{Lines of code changed}
To get a rough estimate of the level of programming effort, we compare the total lines of code before and after incorporating OpenACC (we ignore blank lines). Lines of code increase by 147, from 48,674 to 48,821; this change includes the addition OpenACC directives and other modifications such as the flux-gathering routine and MPI packing/unpacking. Table \ref{tab:loc2} shows the breakdown of the specific OpenACC directives used. 

\begin{table}[h!]
    \centering
    \begin{tabular}{l l l}
    Directive & Description & \# of lines \\
    \hline
      \verb|!$acc parallel|  & Loop parallelism & 29 \\
       \verb|!$acc routine|  & Convert of CPU to GPU routines & 15 \\
       \verb|!$acc loop seq| & Serialize loops & 27 \\
       \verb|!$acc atomic| & Atomic updates & 15 \\
       \verb|!$acc wait| & Barrier & 8 \\
       Continuation lines & & 4 \\
       \hline
       Total & & 98 \\
         \hline
    \end{tabular}
    \caption{Line count of OpenACC directives.}
    \label{tab:loc2}
\end{table}

\section{Conclusion} \label{sec:conclusion}
We have ported DG-SWEM to NVIDIA GPUs using OpenACC Fortran  in a basic manner by exploiting the data parallelism inherent in the explicit-time DG method through element and edge computations . We tested the code on multiple nodes on the NVIDIA Grace Hopper system using a large flooding test case and observed significant speedups over the original CPU implementation. 
When combined with the CUDA Unified Memory System, the porting requires very few code modifications and can be maintained in the same code base; this is a marked improvement from our previous porting framework \cite{DuChene2011-ll}. Finally, we note that there are potential performance issues that can be further investigated on top of this basic parallelization. First, a more detailed  benchmark on multi-node scaling can be done to observe the code's relative performance, and whether further optimizations like overlapping computation and communication will be beneficial. Second, we can explore alternative ways to structure edge integration to exploit GPU shared memory and improve data locality among neighboring edges.

\section{Data Availability}
Some or all data, models, or code that support the findings of this study are available from the corresponding author upon reasonable request.
\begin{itemize}
\item Flood datasets used
    \item DG-SWEM code - \url{https://github.com/foci/dgswem}
\end{itemize}

\section{Acknowledgements}
This work has been supported by the United States National Science Foundation NSF PREEVENTS Track 2 Program, under NSF Grant Numbers 1855047.
This material is also based on work supported by the US Department of Homeland Security under Grant No.2015-ST-061-ND0001-01. The views and conclusions contained in this document are those of the authors and should not be interpreted as necessarily representing the official policies, either expressed or implied, of the US Department of Homeland Security.

The authors also would like to gratefully acknowledge the use of the "ADCIRC" allocation on the Vista supercomputer, and ``ADCIRC", ``DMS23001", and ``DMS21031" allocations on the Frontera supercomputer at the Texas Advanced Computing Center at the University of Texas at Austin. 

\newpage
\bibliography{references}
\newpage
\listoffigures
\listoftables
\listofalgorithms
\end{document}